\documentclass{iaus}
\usepackage{graphics}

\title[Radiative model for eccentric transiting planets] %% give here short title %%
{A time-dependent radiative model for the atmosphere of the eccentric transiting planets}

\author[Nicolas Iro \& Drake Deming]   %% give here short author list %%
{Nicolas Iro$^{1,2}$
 \and Drake Deming$^1$}

\affiliation{$^1$NASA GSFC Planetary Systems Laboratory code 693.0,\\
		     Greenbelt, MD 20770, USA \\ email: {\tt nicolas.iro@nasa.gov ; leo.d.deming@nasa.gov} \\[\affilskip]
$^2$LESIA Observatoire de Paris-Meudon, \\ 92190 Meudon Cedex, FRANCE }

\pubyear{2008}
\volume{253}  %% insert here IAU Symposium No.
\pagerange{}
\date{?? and in revised form ??}
\jname{Transiting Planets}
\editors{}

\newcommand{\rsol}{R$_\odot$}
\newcommand{\rj}{R$_{\rm J}$}
\newcommand{\mj}{M$_{\rm J}$}

\usepackage{aas_macros} %%% inclut les abreviation ADS dans la biblio

\begin{document}

\maketitle

\begin{abstract}
We present a time-dependent radiative model for the atmosphere of the
transiting planets that take into account the eccentricity of their orbit.
We investigate the temporal temperature and flux variations due to the
planet-star distance variability.
We will also discuss observational aspects with {\it Spitzer} measurements.
%% add here a maximum of 10 keywords, to be taken form the file <Keywords.txt>
\end{abstract}

\firstsection % if your document starts with a section,
              % remove some space above using this command.
\section{Introduction}
{\underline{\it Why taking into account the eccentricity}}?
The major theoretical models of atmospheric structure of exoplanets consider
the planet only into two conditions:
\begin{itemize}
\item{} Either the flux received by the planet coming from the star is averaged over an hemisphere of the planet in a static case ;
\item{} or some sort of redistribution day to night of the incoming flux is included.
\end{itemize} 
In none of this cases the fact that the flux actually varies with time is taken into account.
Even in the case of the less eccentric planet, WASP-10 b, with a low eccentricity of 0.057 (\cite[Christian et al.  2008]{Christian08}) this represents a variation $\sim$~25\% in the received flux between periastron and apoastron

We focused our study to two planets:
HD~17156b, the most eccentric transiting planet, 
and even if HD~80606b is not transiting, we choose this planet with the largest eccentricity to show the extreme case.
Tab.~1 summarize the parameters of these two systems.

\begin{table}
  \begin{center}
  \caption{Planetary and stellar parameters}
  \label{tab1}
 {\scriptsize
\begin{tabular}{l|l||c||c}
\hline
&Parameters&HD 17156b &HD80606~b\\
\hline
Orbit%&&&&&&&&\\
     & $e$   & 0.67  & 0.93 \\
     & $a$
       [AU]  & 0.159 & 0.432 \\
     & $d_{\rm min}$
       [AU]  & 0.052 &0.030 \\
     & $d_{\rm max}$
       [AU]  &0.266  & 0.834\\
     & $P$ [days]  &21.22 & 111.78 \\
     & $\omega$
       [$^\circ$]& 121 &300\\
\hline
Planet %&&&&&&&&\\
     & Radius %\tablenotemark{a}
       [\rj] & 0.96 & 1.1$^{\rm a}$\\

     & Mass %\tablenotemark{b}
       [\mj] & 3.11 & 5$^{\rm b}$ \\
%     & \fint [] & & & & & & \\
     & $P_{\rm spin}$
       [days] &    3.8 &1.93\\
\hline
Star %&&&&&&&&\\
     & Type  & G0V   &  G5\\
     & Radius
      [\rsol]&1.47 &1.05  \\
     &[Fe/H] & 0.24  & 0.43\\
\hline

Refs.&        &  \cite{Gillon07}& \cite{LL08}\\
\hline
\end{tabular}\label{tab:param}

  }
 \end{center}
\vspace{1mm}
 \scriptsize{
 {\it Notes:}\\
$^{\rm a}$ This planet is not transiting. The value of the radius has been predicted by (4) ; \\
 $^{\rm b}$We adopted this value of the mass (inclination unknown).
}

\end{table}

        \section{Model description}
In order to model the planets atmosphere, we use a line-by-line radiative transfer code which is described in details in \cite{Iro05}.

The equation we have to solve is:
\begin{equation}
\frac{d T}{d t}= 
\underbrace{
-\frac{m g}{C_{\rm p}}\frac{d F_\star}{d p}
}_{\mbox{\rm Heating rate}}
 - \underbrace{
\left(-\frac{m g}{C_{\rm p}}\frac{d F_{\rm IR}}{d p}\right)
}_{\mbox{\rm Cooling rate}}\label{equ1}
\end{equation}
where $T$ is the temperature, $t$ time and we divided the flux $F$ into 2 quantities:
the thermal flux emitted by the planet (upward-downward) $F_{\rm IR}$ and the stellar net flux (downward-upward)$F_\star$, so that we have $F=F_{\rm IR} - F_\star$.

We take advantage of the fact that the time is explicitly included in the equation by introducing a variation of the insolation at each time step.
This variation can be due to the non-constant planet-to-star distance during the eccentric orbit and/or to the rotation of the atmosphere since the eccentric planets have a remnant spin.
They are indeed supposed to be in pseudo-rotation.
We calculated the rotation period using the prescription of \cite{Hut81} (see Tab.~1).

           \subsection{Method}

Here is the scheme of our calculations:

 \subsubsection{Static calculations}

The first thing we do is calculate the thermal structure of a planet with the planet characteristics located at the semi major axis.

We do so iteratively. That means that 
           \begin{enumerate}
\item for an initial thermal profile we calculate the heating and cooling rates
\item  then make them converge by modifying the thermal profile according to Eq.~\ref{equ1} setting the left-hand term to 0 ({\bf thermal equilibrium})
\item if the profile is different by the end of process ({\it b}) ; we begin process ({\it a}) with the new profile ; if it is the same that means that we have reached the static solution.
           \end{enumerate}

           \subsubsection{Orbit calculation}

We then calculate for each time the actual planet-star distance and take the distance factor into account to correct for the right insolation. This is a modified version of point ({\it b}) from above.

\begin{figure}
\centering
%\resizebox{6.5cm}{!}{\includegraphics{2ecc_HD80606_B_more.eps} }
%\resizebox{6.5cm}{!}{\includegraphics{2ecc_HD17156_B_more.eps} }
\resizebox{13cm}{!}{\includegraphics{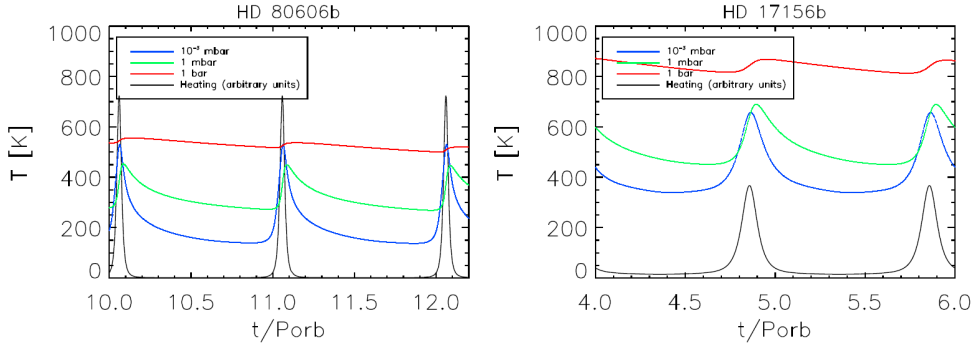} }
\caption[]{Temperature of the planet atmospheres at selected pressure levels as a function of time\label{fig:res1} for the condition where the flux received by the planet is averaged over its sphere.
We let the calculation make several orbital revolution in order to reach a periodic solutions (note the different scales).}
\end{figure}

\begin{figure}
\centering
%\resizebox{6.5cm}{!}{\includegraphics{2becc_HD80606_B_more.eps} }
%\resizebox{6.5cm}{!}{\includegraphics{2becc_HD17156_B_more.eps} }
\resizebox{13cm}{!}{\includegraphics{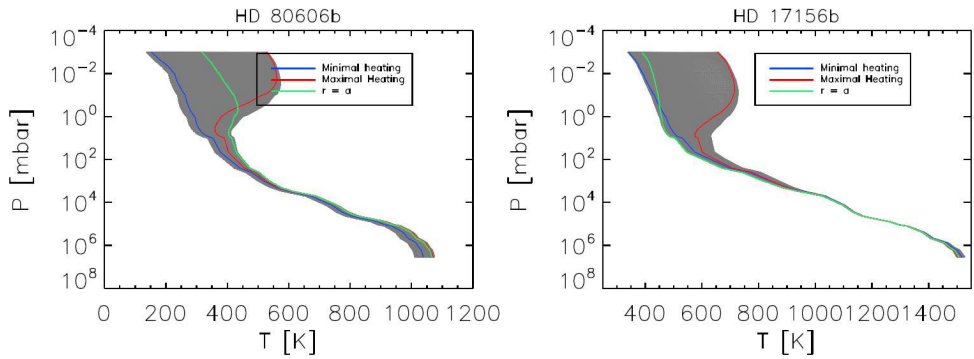} }
\caption[]{Temperature Structures of the atmospheres at selected orbit positions ({\it lines}) and for every times ({\it shaded area}) (note the different scales).\label{fig:res2}}
\end{figure}

We obtain a thermal profile as well as the spectra for each time step.
The results of this step is shown in Fig.~\ref{fig:res1} and Fig.~\ref{fig:res2} for the thermal structures and in Fig.~\ref{fig:res3} for the fluxes.

          \subsubsection{Rotation calculation}

Eventually we add the rotation of the planet.
These eccentric planets are indeed supposed to not be completely synchronized.
But we choose to implement the rotation along with a multi-latitude / multi-longitude approach (and a parallelization of the code).
This step is still on the way (see the Perspectives section for further details). 
          \subsection{Limitations}

Here is some simplifications that we have done:
           \begin{itemize}
\item The chemical composition is not changed as the thermal profile does.
The species are calculated at the semi major axis distance.
\item  In the same way, the heating rates and cooling rates are for now only calculated at the semi major axis.
Then only the varying {\bf stellar heating} is accounted for.
\item we have not included any clouds due to the uncertainties on the formation processes.
           \end{itemize}

\begin{figure}
\centering
%\resizebox{6.5cm}{!}{\includegraphics{sp2_HD80606_B_more.eps} }
%\resizebox{6.5cm}{!}{\includegraphics{sp2_HD17156_B_more.eps} }
\resizebox{13cm}{!}{\includegraphics{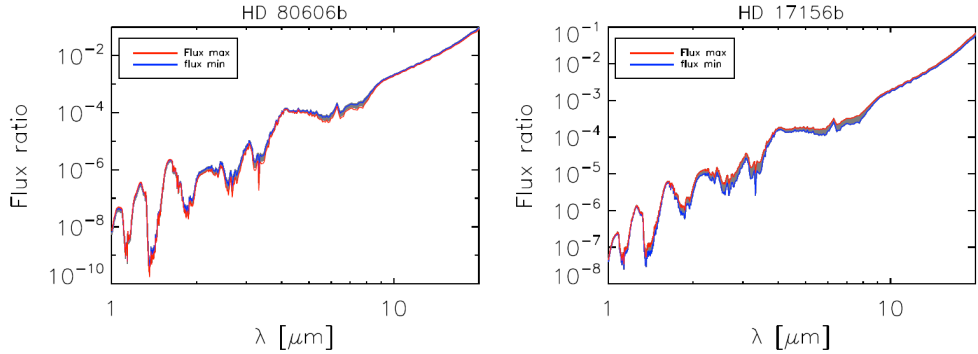} }
\caption[]{Ratio of the planet flux by the star flux as a function of wavelength\label{fig:res3} at two positions ({\it lines}) and for every times ({\it shaded area}).}
\end{figure}

            \subsection{Discussion}

	   Fig.~\ref{fig:res3} compares the ratio of the planet flux over the star flux as a function of wavelength.
We want to draw your attention on the fact that before comparing directly with {\it Spitzer} observations, we have to take into account the phase effects.
Here, we have an average over the planet.
During secondary eclipses we observe the illuminated hemisphere, this leads to very different values of the flux \ldots
In the next development of the code (see the Perspective section), this will be investigated.

         \section{Perspectives}

             \begin{itemize}
         
             \item We have benefited the parallelization capabilities of our code to include a different heating pattern depending on the longitude and latitude. By adding also the planet rotation with the samemethod as in \cite{Iro05}, we will be able to construct a thermal map of the planet as a function of time.
By integrating the view corresponding to the observation, we will also be able to calculate a full light curve in any wavelength band.
             \item We then will also include a calculation of the heating rates and cooling rates for each planet point as well as the corresponding thermal profiles and chemical composition.
             \item Finally, we will also calculate the above quantity as a function of time.
This will require a huge increase of the calculation speed. The use of the correlated k-coefficient method (opposed to line-by-line) will probably be necessary.
             
             \end{itemize}

\section{Conclusion}

We presented a still under development time-dependent radiative model for the atmosphere of the
transiting planets.
This model takes into account the eccentricity of their orbit and in its next version will be able to completely map the planet, including the rotation effects on the insolation.

The preliminary results shows already that an eccentric planet in under heating conditions is very different that if it were sitting at its semi major axis, even for a low eccentricity even if several steps need to be completed before reaching a fully self-consistent model.

\end{document}